\documentstyle[12pt]{article}

\def\refs{\leftskip=.3truein\parindent=-.3truein}
\def\unrefs{\leftskip=0.0truein\parindent=20pt}

\title {Look Back Time, the Age of the Universe, and the Case for
  a Positive Cosmological Constant\thanks{To be published in
 the {\em Journal of the Royal Astronomical Society of Canada},
 August, 1993 issue.}}

\author {Kevin Krisciunas}
\date { }

\begin{document}
\maketitle

\vspace {-1 cm}

\begin {center}
     Joint Astronomy Centre \\
     660 N. A'ohoku Place \\
     University Park \\
     Hilo, Hawaii 96720 USA
\end {center}

\vspace{5 mm}

\begin{abstract}

We present explicit expressions for the calculation of
cosmological look back time, for zero cosmological constant
and arbitrary density parameter $\Omega$, which, in the limit as
redshift becomes infinite, give the age of the universe. The
case for non-zero cosmological constant is most easily solved via numerical
integration.  The most distant objects presently known
(approaching redshift z = 5) have implied ages of $\approx$ 1-2 Gyr
after the the Big Bang.  The range of such age is narrow, in
spite of a variety of cosmological models one might choose.
We give a graphical representation of a variety of
cosmological models and show that a wide range of Hubble
constants and values of the age and density of the universe
compatible with modern studies are consistent with adoption of
a positive cosmological constant.
\end{abstract}

Assuming the correctness of the standard Big Bang
scenario (Peebles {\em et al.} 1991), the redshift (z) of a distant
galaxy or quasar can be related (Longair 1984, Eq. 15.19) to
the cosmic scale factor of the universe (R), as follows:

\begin{equation}
R = \frac{1}{1 + z} \;.
\end{equation}

\vspace{5mm}

\parindent=0pt

R = 1 at the present epoch.  An object at z = 1 emitted its
light when the universe was half its present scale (R = 0.5).
{\em How long ago} the light was emitted (the look back time
$\tau$ ) depends on the dynamics of the universe.

\parindent=20pt

The look back time (following Longair 1984, Eq. 15.47), and assuming
for now a zero cosmological constant, is:

\begin{equation}
\tau = \frac{1}{H_o}\int_{0}^{z}\frac{dz'}
          {(1 + z')^2 (\Omega z' + 1)^{1/2}} \;.
\end{equation}

\vspace{5mm}

\parindent=0pt

Here the density parameter $\Omega = \frac {\rho_o}{\rho_c}$
is the ratio of the
density of the universe to the critical density

\begin{equation}
\rho_c = \frac{3 H_o^2}{8 \pi G} = 1.88 \times 10^{-29}
    \left[\frac{H_o}{100}\right]^2 \; g/cm^3 \;.
\end{equation}

\vspace{5mm}

If the vacuum energy density of the universe is zero (i.e. if
Einstein's cosmological constant $\Lambda = 0$), $\Omega < 1$ implies a
universe with negative curvature which will expand forever;
$\Omega > 1$ implies a universe with positive curvature which will
eventually recollapse; and $\Omega = 1$ (the ``Einstein-de Sitter
universe") implies a universe with flat geometry which will
expand forever, but which will eventually reach zero
expansion rate.

\parindent=20pt

In Equations 2 and 3 above $\rm {H_o}$ is the Hubble constant,
measured in km/sec/Mpc.  Note that it is measured in terms of
distance per unit time per some other unit of distance, or
$\rm (time)^{-1}$.  If we denote h = $\rm {H_o}$ / (100 km/sec/Mpc), then we
speak of a ``Hubble time" (in units of 100 km/sec/Mpc) as
follows:

\begin{equation}
T_H = H_o^{-1} = 9.778 \: h^{-1} \; (Gyr)\; .
\end{equation}

\vspace{5mm}

\parindent=0pt

For $\rm {H_o}$ = 50 km/sec/Mpc, h = 0.5 and the Hubble time is 19.6
billion years.  It is important to note that the age of the
universe can be less than or greater than the Hubble time,
depending on the value of $\Omega$ and whether or not Einstein's
cosmological constant ($\Lambda$) is zero.  But if $\Lambda  = 0$, the age of
the universe is necessarily less than the Hubble time,
because of the gravitational ``braking effect" of the matter
in the universe.  (In that case the present expansion rate of
the universe is necessarily less than it was in the past.)

\parindent=20pt

For an empty universe ($\Omega  = 0$) Equation 2 has the simple
solution:

\begin{equation}
\tau = T_H \left( \frac{z}{1+z} \right) \;.
\end{equation}

\vspace{5mm}

For $\Omega = 1$,  Equation 2 has another simple solution:

\begin{equation}
\tau = \frac{2}{3} \: T_H \left[ 1 - \frac{1}{(1 + z)^{3/2}} \right] \; .
\end{equation}

\vspace{5mm}

For $\Omega > 1$, we can solve Equation 2 with the help of Gradshteyn
and Ryzhik (1965):

\begin {displaymath}
\frac{\tau}{T_H} = \frac {(1 + \Omega z)^{1/2}}{(\Omega - 1)(1+z)} +
\frac{\Omega}{(\Omega - 1)^{3/2}} \: \rm{tan}^{-1} \left [ \left (
   \frac {1 + \Omega z}{\Omega - 1} \right) ^{1/2} \right ]
\end {displaymath}

\begin{equation}
   - \frac{1}{(\Omega - 1)} -
   \frac{\Omega}{(\Omega - 1)^{3/2}} \: \rm{tan}^{-1} \left[ \left (
   \frac {1} {\Omega - 1} \right) ^{1/2} \right] \; .
\end{equation}

\vspace{5mm}

\parindent=0pt

And for $0 < \Omega < 1$,  it can be shown (Gradshteyn and Ryzhik 1965)
that the look back time is:

\begin {displaymath}
\frac{\tau}{T_H} = \frac {- (1 + \Omega z)^{1/2}}{(1- \Omega)(1+z)} -
\frac{\Omega}{2 (1 - \Omega)^{3/2}} \: \rm{ln} \left [
\frac {(1 + \Omega z)^{1/2} - (1 - \Omega)^{1/2}}
      {(1 + \Omega z)^{1/2} + (1 - \Omega)^{1/2}} \right]
\end {displaymath}

\begin{equation}
   + \frac{1}{(1 - \Omega)} +
   \frac{\Omega}{2 (1 - \Omega)^{3/2}} \: \rm{ln} \left[
   \frac {1 - (1 - \Omega)^{1/2}} {1 + (1 - \Omega)^{1/2}} \right] \; .
\end{equation}

\vspace{5mm}

While Equations 5 and 6 are often used to describe look back
time, Equation 8 is a much more realistic one to use, since
many studies imply $\Omega \approx 0.1$ (e.g. Ford {\em et al.} 1981,
Press and Davis 1982).

\parindent=20pt

Other solutions mathematically equivalent to Equations 7 and 8 have
appeared in the literature.  For the $\Omega > 1$ case it is easy to
show that Equation 23 of Sandage's (1961b) parametric method
is equal to Equation 3.24 {\em minus} Equation 3.22 of Kolb and Turner (1990).
For the $0 < \Omega < 1$ case it
is not difficult to show that Sandage's Equation 24 equals Kolb and Turner's
Equation 3.25 {\em minus} Equation 3.23, providing one uses the identity:

\begin {displaymath}
\rm{cosh}^{-1}\theta = \rm{sinh}^{-1} (\sqrt{\theta^2 - 1} ) \; .
\end {displaymath}

\vspace{5mm}

\parindent=0pt

Then, using trigonometric identities for cos(x-y), cos(x/2) and
sin(x/2) it is possible to show that Kolb and Turner's Equations 3.24
{\em minus} 3.22 equals our Equation 7 above.  (Only the {\em sum} of
the second and
fourth terms of our Equation 7 equals the {\em sum} of two of the terms
from Kolb and Turner.)  For the $0 < \Omega < 1$ case it can be shown
that our Equation 8 equals Kolb and Turner's Equation 3.25 {\em minus}
Equation 3.23 providing we use the identity:

\begin{displaymath}
\rm{cosh}^{-1} \theta = \rm{ln} [ \theta + (\theta^2 - 1)^{1/2} ] \; .
\end {displaymath}

\vspace{5mm}

Conveniently, our Equation 8 works out term by term with the solution
of Kolb and Turner.  Finally, we note that Schmidt and Green (1983) give
an expression equivalent to our Equation 8 divided by Equation 15 (below).

\parindent=20pt

For $\Lambda \neq 0$, following Caroll {\em et al.} (1992, Equation 16)
or Krisciunas (1993, Equation 27), Equation 2 becomes

\begin{equation}
\frac {\tau}{T_H} =  \int_0^z \frac {dz'}
{(1 + z')^2 \left\{ \Omega z' + 1 - \lambda \left [
 \frac {2 z' + (z')^2}{(1 + z')^2} \right] \right\}^{1/2} } \; .
\end{equation}

\vspace{5mm}

\parindent=0pt

where

\begin{equation}
\lambda = \frac {\Lambda c^2}{3 H_o^2}
\end{equation}

\vspace{5mm}

\parindent=0pt

is the ``reduced" (dimensionless) cosmological constant.
Defined this way $1/\sqrt{\Lambda}$ has units of length.  This is the
``length scale over which the gravitational effects of a
nonzero vacuum energy density would have an obvious and
highly visible effect on the geometry of space and time"
(Abbott 1988).  In general Equation 9 is most easily solved by means
of numerical integration.

\parindent=20pt

In Figure 1 we show the look back time {\em vs.} redshift for
four scenarios: $\rm {H_o} = 50$ and $\Omega$ = 0.0, 0.3, and 1.0 (with
$\lambda = 0$); and $\rm {H_o} = 80$, $\Omega  = 0.12$,
$\lambda = 0.88$.  The advantages of the last model will become
clear shortly.

In Figure 2 we show the look back time, measured in
terms of the age of the universe (using Equations 14, 15, and
17 below).

According to the inflationary scenario (Narlikar and
Padmanabhan 1991), the universe should have flat geometry, a
condition satisfied by:

\begin{equation}
\Omega + \lambda = 1 \; .
\end{equation}

\vspace{5mm}

\parindent=0pt

So either: 1) $\Omega = 1$ (and we cannot find the missing mass even
from the dynamics of clusters of galaxies); 2) $\Omega < 1$ and
$\lambda  = 1 - \Omega$; or 3) the inflationary paradigm is incorrect.

\parindent=20pt

In the limit as $z \rightarrow \infty$ for Equations 5 through 8, we
obtain expressions for the age of the universe ($\rm {T_o}$), assuming
$\lambda = 0$:

\begin{equation}
T_o = T_H \; . \;\;\;\; (\Omega = 0)
\end{equation}

\vspace{5mm}

\begin{equation}
T_o = \frac{2}{3} T_H \; . \;\;\;\; (\Omega = 1)
\end{equation}

\vspace{5mm}

\begin{equation}
\frac {T_o}{T_H} = \frac {-1}{(\Omega - 1)} +
   \frac {\Omega}{(\Omega -1)^{3/2}} \rm{sin}^{-1} \left[
   \left( \frac {\Omega - 1}{\Omega}\right)^{1/2} \right] \; .
   \;\;\;\; (\Omega > 1)
\end{equation}

\vspace{5mm}

\begin{equation}
\frac {T_o}{T_H} = \frac {1}{(1 - \Omega)} -
   \frac {\Omega}{2 (1- \Omega)^{3/2}} \rm{ln} \left[
   \frac{2-\Omega}{\Omega} + \frac{2 (1-\Omega)^{1/2}}{\Omega}
   \right] \;. \;\;\;\; (0 < \Omega < 1)
\end{equation}

\vspace{5mm}

\parindent=0pt

Our Equation 14 is mathematically equivalent to Equation 61 of
Sandage (1961a), given $q_o = \Omega / 2$ for $\lambda = 0$.
Our Equation 15 is equal to Equation 65 of Sandage (1961a) given the
same definition of $q_o$.

\parindent=20pt

It should be noted that Equations 14 and 15 are more
easily derived by integrating an expression for the Hubble
constant divided by the rate of change of the cosmic scale
factor over the age of the universe:

\begin{equation}
\frac{T_o}{T_H} = \int_0^{T_o} \frac {H_o}{(dR/dt)} =
   \int_0^1  \frac{dR} {\left[ \frac {\Omega}{R} + 1 -
  \Omega + \lambda (R^2 - 1) \right]^{1/2}} \; ,
\end{equation}

\vspace{5mm}

\parindent=0pt

as is done in Krisciunas (1993, Appendix A).

\parindent=20pt

For the special case of $0 < \Omega < 1$ and $\Omega + \lambda = 1$,
Equation 16 can be solved by making the substitution:

\begin {displaymath}
R = \left( \frac {\Omega}{1-\Omega} \right) ^{1/3} \:
\rm{sinh}^{2/3} \theta
\end {displaymath}

\vspace{5mm}

\parindent=0pt

and by using the identity

\begin {displaymath}
\rm{sinh}^{-1} \theta = \rm{ln} [ \theta + (\theta^2 + 1)^{1/2} ] \; .
\end {displaymath}

\vspace{5mm}

The solution is:

\begin{equation}
\frac{T_o}{T_H} = \frac{2}{3}\: \frac {1} {(1 - \Omega)^{1/2}} \: \rm{ln}
  \left[ \frac {1 + (1 - \Omega)^{1/2}} {\Omega^{1/2}} \right] \; .
 \;\;\;\; (0 < \Omega < 1; \: \Omega + \lambda = 1)
\end{equation}

\vspace{5mm}

This is equivalent to Equation 3.32 of Kolb and Turner (1990).

\parindent=20pt

Assuming values for $\Omega$ and $\lambda$ allows us to calculate the
age of the universe in Hubble times.  Assuming $\rm {H_o}$ allows us
to calculate the Hubble time in billions of years.  In Figure
3, assuming $\lambda = 0$, we show various loci of points
corresponding to a range of Hubble constants, values of the
age of the universe, and density scale factors.

Now a common idea associated with the construction of
the new 8-10 meter class telescopes is that they will ``allow
us to see 13 billion years into the past, about 1 to 2
billion years after the Big Bang."  It turns out that only
the second half of that statement is necessarily correct.

The largest observed quasar redshift is presently 4.897
(Schneider {\em et al.} 1991), and the largest observed galaxy
redshift is 3.8 (Chambers {\em et al.} 1990).  Let us consider the
look back time of objects of redshift 5, regarded by many to
be the beginning of the era of galaxy formation (Ellis 1987).
Using Equations 6, 8, 13, 15 and 17, and numerical solutions
to 9, we give a number of examples in Table 1.  For all these
cases the age of a z = 5 object, reckoned since the Big Bang,
is about 1 to 2 billion years, even though the look back time
(depending on the model) ranges from 9.68 to 15.49 Gyr.
Because of uncertainties in choosing the right model, we do
not really have an accurate handle on the {\em look back time} of a
z = 5 object (because the range of derived values of the {\em age}
of the universe), but we can say with some certainty that
that light originated 1 to 2 Gyr after the Big Bang.

Let us now consider the implications of the loci of
points in Figure 3.  Following Fowler (1987), we agree that
our understanding of single star evolution is on a
reasonably firm footing, though it should be pointed out that
changes in the assumed abundances of certain elements (e.g.
oxygen) have led to significant revisions in the ages of the
oldest known stars in our galaxy.  We could have extended the
loci of points in Figure 3 further to the right -- the universe
could be older than 18 Gyr. That would further strengthen
the idea that the Hubble constant must have a ``small" value or that
$\lambda > 0$ (see below).

One often estimates the age of the universe by
deriving the age of the oldest stars in our galaxy and adding
a ``sensible" incubation time for our galaxy.  Let us suppose
for a moment that the age of the oldest stars in our galaxy is
equal to that value derived in the careful study of the globular
cluster 47 Tucanae by
Hesser {\em et al.} (1987), who find an age of 13.5 $\pm$ 0.5 Gyr
(internal error; $\pm$ 2.0 Gyr external error).  The formal error
bars should not be taken too seriously, but we can take a
representative age-since-the-Big-Bang of 1.5 Gyr from Table
1, and add it to the age of 47 Tucanae, 13.5 Gyr,
giving an age of the universe, $\rm {T_o}$, of 15.0 Gyr.

 What is the most valid value to adopt for the Hubble
constant?  This is a subject too complex to investigate in
any detail here, but values of $\rm {H_o}$ based on the infrared
Tully-Fisher relation cluster around 80 km/sec/Mpc.  Okamura
and Fukugita (1992) give a graphical summary.  Two recent
reviews (Jacoby {\em et al.} 1992; van den Bergh 1992) give $\rm {H_o}$ = 80
$\pm$ 11 and $\rm {H_o}$ = 76 $\pm$ 9, respectively.    The Sandage-Tammann
value of about 50 km/sec/Mpc also has its strong advocates
(though they presently seem to be the minority).

If $\rm {T_o}$ = 15.0 Gyr and $\rm {H_o}$ = 80 km/sec/Mpc, an inspection
of Figure 3 immediately indicates that the density of the
universe must by less than zero!  For a zero cosmological
constant, if $\rm {T_o}$ = 15.0 Gyr, $\rm {H_o} \leq 65$.  If $\rm {H_o}$ = 80,
then
$\rm {T_o} \leq 12.2$ Gyr.  So either the ``most sensible" value of the Hubble
constant is wrong; {\em or} we need to revise the models from which
we derive the ages of the oldest stars;  {\em or} the cosmological
constant might be non-zero.  Now, this is not a new idea, but
the simplicity of Figure 3 has the advantage that it can be
easily understood by non-cosmologists.  van den Bergh (1992)
points out that studies of globular cluster ages give a range
of 12 to 17 Gyr, so our adoption of 13.5 Gyr and a galaxy
incubation time of 1.5 Gyr may not be intellectually honest.
But as long as determinations of the Hubble constant give $\rm {H_o}
\approx 80$, we must take seriously the idea of $\lambda > 0$.

A positive cosmological constant is the same as
attributing ``repulsive" force to the vacuum.  As the universe
expands, this repulsive force becomes stronger and stronger, as there
there is more space within which it works.  To illustrate the
extent of the effect, consider the basic idea of Equation 16:

\begin{equation}
  \frac {(dR/dt)} {H_o} =
  \left[ \frac {\Omega}{R} + 1 - \Omega + \lambda (R^2 - 1) \right]^{1/2}
  \; .
\end{equation}

\vspace{5mm}

\parindent=0pt

In Figure 4 we graph this function for $\Omega = 2.0$ and various
small positive values of $\lambda$.  One can see that $\lambda$
only need be as large as $\approx +0.042$ for the rate of change of R
to be positive always.  In other words, the universe can expand
forever, even in cases where $\Omega + \lambda$ is significantly
greater than 1.

\parindent=20pt

The effect of a positive cosmological constant is
to take the loci of points in Figure 3 and pull them up in
the diagram (see Figure 5).  Scanning through the examples in
Table 1, if $\Omega = 0.1186$, and $\lambda  = 1 - \Omega$,
then we find that $\rm {T_o} = 15.0$ Gyr if $\rm {H_o} = 80$ km/sec/Mpc.
This satisfies the
theoreticians who advocate the verisimilitude of the
inflationary scenario, while also allowing our understanding
of stellar evolution and the values of $\rm {H_o}$ based on the
infrared Tully-Fisher relation to stand as valid.  But
perhaps this is only a modern example of the ancient Greek
method of ``saving the phenomenon".\footnote{For example,the
ancient Greeks asserted that a planet's motion
must be {\em uniform} and {\em circular}.  The planet moves
uniformly on its circular deferent around the Earth.  But, because
of the observed effect of retrograde motion, they ``saved the
phenomenon" of uniform, circular motion by inventing the idea of
an epicycle, a smaller circle that turned at a faster rate than
the deferent and whose center rode around on the deferent.  Similarly,
they invented {\em eccentric} deferents and {\em equants} to match
the observed planetary positions better with the ephemerides.  (See
Krisciunas 1988.)
As a modern
example of ``saving the phenomenon", to explain the advance of the
perihelion of the orbit of Mercury and retain Newtonian gravity,
astronomers postulated the existence of the planet Vulcan.  This
certainly worked for the anomalies of the motion of Uranus, which led to
the discovery of Neptune.  However, for Mercury's motion what was
needed was a new theory of gravity, namely General Relativity.
We use these two examples to point out that if {\em ad hoc} hypotheses
such as the cosmological constant are ``required", perhaps the
solution resides in ``proper" correction of some of the other data,
or a significant revision of the theoretical underpinnings.}

Allowing $\rm {T_o}$ to range from 14.5 to 15.5 Gyr, allowing $\rm {H_o}$
to range from 70 to 90 km/sec/Mpc, and assuming that $\Omega + \lambda = 1$,
gives $0.06 < \Omega < 0.23$. Such a range of $\Omega$ is in accord with
studies of large scale structures in the universe (e.g. Ford
{\em et al.} 1981, Press and Davis 1982), as well as Big Bang
nucleosynthesis studies (Boesgaard and Steigman 1985).  Also,
$\lambda \approx$  0.8-0.9 does not contradict a recent constraint on the
cosmological constant derived from the study of gravitational
lens galaxies (Kochanek 1992), that $\lambda < 0.9$.

In choosing a particular cosmological model one must
pick {\em families} of parameters ($\rm {T_o}$, $\rm {H_o}$,
$\Omega$, $\lambda$) that satisfy
particular mathematical relationships.  Not all combinations
are allowed.  Sensible values of $\rm {T_o}$ and $\rm {H_o}$ point to a
positive value for the cosmological constant. (See Tayler
1986 for a similar discussion.) However, various theoretical
discussions (e.g. Weinberg 1989) stipulate that the
cosmological constant should be identically zero.

\newpage

\begin{center}
{\bf References}
\end{center}

\refs

Abbott, L., 1988, Sci. Amer., 258, No. 5 (May issue), 106

Boesgaard, A. M.,  and Steigman, G., 1985, Ann. Rev. A \& A,
   23, 319

Caroll, S. M., Press, W. H., and Turner, E. L., 1992, Ann. Rev.
A \& A, 30, 499

Chambers, K. C., Miley, G. K., and van Breugel, W. J. M.,
  1990, ApJ, 363, 21

Ellis, R., 1987, in {\em High Redshift and Primeval Galaxies}, ed.
   J. Bergeron {\em et al.}, (Gif Sur Yvette, France), p. 3

Ford, H. C., Harms, R. J., Ciardullo, R., and Bartko, F.,
  1981, ApJ, 245, L53

Fowler, W. A., 1987, QJRAS, 28, 87

Gradshteyn, I. S.,  and Ryzhik, I. M., 1965,  {\em Table of
  Integrals, Series, and Products}, (Academic Press, New
  York), p. 78

Hesser, J. E., Harris, W. E., VandenBerg, D. A., Allwright,
  J. W. B., Shott P., and Stetson, P., 1987, PASP, 99, 739

Jacoby, G. H., Branch, D., Ciardullo, R., Davies, R. L.,
  Harris, W. E., Pierce, M. J., Pritchet, C. J., Tonry,
  J. L., and Welch, D. L., 1992, PASP, 104, 599

Kochanek, C. S., 1992, ApJ, 384, 1

Kolb, E. W., and Turner, M. S., 1990, {\em The Early Universe},
  (Addison-Wesley, Redwood City, California)

Krisciunas, K., 1988, {\em Astronomical Centers of the World},
  (Cambridge Univ. Press, Cambridge and New York), chapter 1

Krisciunas, K., 1993,  ``Fundamental cosmological
  parameters," in {\em Encyclopedia of Cosmology}, ed. N. S.
  Hetherington,  (Garland, New York and London), p. 218

Longair, M. S., 1984,  {\em Theoretical Concepts in Physics},
  (Cambridge Univ. Press, Cambridge and New York), chapter 15

Narlikar, J. V.,  and Padmanabhan, T., 1991, Ann. Rev. A \& A,
  29, 325

Okamura, S., and Fukugita, M. 199[2].  ``istance to the Coma
  cluster and the value of $\rm {H_o}$," in {\em Primordial
  Nucleosynthesis and Evolution of the Early Universe}, in press

Peebles, P. J. E., Schramm, D. N., Turner, E. L., and
  Kron, R. G., 1991, Nature, 352, 769

Press, W. H., and Davis, M., 1982, ApJ, 259, 449

Sandage, A., 1961a, ApJ, 133, 355

Sandage, A., 1961b, ApJ, 134, 916

Schmidt, M, and Green, R. F., 1983, ApJ, 269, 352

Schneider, D. P., Schmidt, M., and Gunn, J. E., 1991, AJ,
  102, 837

van den Bergh, S., 1992, PASP, 104, 861

Weinberg, S., 1989, Rev. Mod. Phys., 61, 1

\unrefs

\newpage

\begin{center}
{\bf Figure Captions}
\end{center}

\parindent=0pt

Fig. 1 - Look back time {\em vs.} redshift for four cosmological
models.  A1, A2, and A3 have $\rm {H_o}$ = 50 km/sec/Mpc, $\lambda = 0$, and
$\Omega$ = 0.0, 0.3, and 1.0, respectively.  Model B1 has $\rm {H_o} = 80$
km/sec/Mpc, $\lambda = 0.88$, $\Omega = 0.12$.

\vspace {5mm}

Fig. 2 - Look back time, measured in terms of the age of the
universe, vs. redshift.  A1, A2, and A3 have $\lambda = 0$, and
$\Omega$ = 0.0, 0.3, and 1.0, respectively.  Model B1 has
$\lambda = 0.88$, $\Omega = 0.12$.

\vspace {5mm}

Fig. 3 - Loci of constant $\Omega$ for a range of Hubble constants
and values of the age of the universe ($\rm {T_o}$).  These models
assume a zero cosmological constant.

\vspace {5mm}

Fig. 4 - The rate of change of the cosmic scale factor (divided by
the present value of the Hubble constant) {\em vs.} the cosmic scale
factor for $\Omega = 2.0$ and various small positive values of the
scaled cosmological constant.
If the curve intersects the X-axis, the model indicates a
universe that reaches a maximum scale, then recollapses.  Otherwise,
the model indicates a universe that expands forever.

\vspace {5mm}

Fig. 5 - Loci of constant $\Omega$ for a range of Hubble constants
and values of the age of the universe ($\rm {T_o}$), assuming the
correctness of the condition stipulated by the inflationary
scenario, $\Omega + \lambda = 1$.

\end{document}